# Representing high throughput expression profiles via perturbation barcodes reveals compound targets

*Tracey Filzen[1], Peter Kutchukian[2], Jeffrey Hermes[1], Jing Li[1], Matthew Tudor[1*]*

[1] Screening & Protein Sciences, Merck Research Laboratories, North Wales, PA

[2] Cheminformatics, Merck Research Laboratories, Boston, MA

* to whom correspondence should be addressed: matthew_tudor@merck.com

## Abstract

High throughput mRNA expression profiling can be used to characterize the response of cell culture models to perturbations such as pharmacologic modulators and genetic perturbations. As profiling campaigns expand in scope, it is important to homogenize, summarize, and analyze the resulting data in a manner that captures significant biological signals in spite of various noise sources such as batch effects and stochastic variation. We used the L1000 platform for large-scale profiling of 978 genes, chosen to be representative of the genome as whole, across thousands of compound treatments. Here, a method is described that uses deep learning techniques to convert the expression changes of the landmark genes into a perturbation barcode that reveals important features of the underlying data, performing better than the raw data in revealing important biological insights. The barcode captures compound structure and target information, in addition to predicting a compound's high throughput screening promiscuity, to a higher degree than the original data measurements, indicating that the approach uncovers underlying factors of the expression data that are otherwise entangled or masked by noise. Furthermore, we demonstrate that visualizations derived from the perturbation barcode can be used to more sensitively assign functions to unknown compounds through a guilt-by-association approach, which we use to predict and experimentally validate the activity of compounds on the MAPK pathway. The demonstrated application of deep metric learning to large-scale chemical genetics projects highlights the utility of this and related approaches to the extraction of insights and testable hypotheses from big, sometimes noisy data.

## Background

Pharmacology is generally explored in a linear and iterative manner, starting from the observation of an activity, that is then optimized for selectivity and potency via various methods, and subsequently tested in preclinical and randomized controlled clinical studies for safety and efficacy. In general, the activity of compounds on targets other than the intended one(s) is limited, though such 'off-target' activities may lead to adverse events. The detection of polypharmacology is desired and could be facilitated by comprehensive profiling of drug candidates on a general phenotyping platform, i.e. a consistent method to broadly assess the phenotypic effects of compound treatment. Such a platform could be used in order to uncover unexpected phenotypic signals beyond the originally identified mechanism-based phenotype.

Phenotypic screening is considered by some to be a return to pharmacology's roots (Swinney and Anthony 2011), and by others a new discipline, that will need to prove itself over the coming decades (Eder, Sedrani et al. 2014). In either case, there is increased interest in applying cell-based pathway or phenotype screens to identify unknown/unexpected targets along with tool compounds that can be used for target validation, and potentially as starting points for drug discovery. One major hurdle in phenotypic screening approach *versus* a target-based approach lies in the identification of the target(s) of molecules that show an activity in cell-based (or organismal) assays. A general phenotyping platform could be used to infer mode of action of unknown compounds based on similarity of induced expression profiles' similarity to those of annotated compounds. Such data can also in some cases be used to propose new indications for known molecules (Lamb, Crawford et al. 2006).

Lastly, a general phenotyping platform will allow one to monitor compounds through their maturation and optimization in order to prioritize series based on selectivity and to quickly identify potential polypharmacology and safety warning signals (Verbist, Klambauer et al.).

We suggest that mRNA is an ideal analyte for a general phenotyping platform. Whereas gene expression changes are often distal to signaling and metabolic pathways that drug discovery aims to modulate, most perturbations of such pathways eventually lead to the nucleus and to transcriptional changes that propagate, amplify, or compensate for the immediate effects of a perturbation. mRNA also has the beneficial property that its measurement is fairly easy to generalize, such that any set of target sequences can be measured quantitatively and in parallel. Thus, a potentially broadly useful general phenotyping platform would quantitate mRNA, be medium to high throughput, be affordable to apply to thousands of samples, and produce highly reproducible data.

The L1000 platform (Peck, Crawford et al. 2006) has the potential to be just such a general phenotyping platform, one that can be used in various stages of drug discovery, including target identification and validation, hit-to-lead, lead optimization, as well as safety assessment and repurposing. 978 genes were selected to be representative of the expression of the remainder of the transcriptome (Duan, Flynn et al. 2014), and the platform is used to capture the transcriptional phenotypes using this reduced set of 'landmark' genes. The high throughput and relatively low cost of the bead array based implementation permits comprehensive application to large numbers of perturbations, be they different compounds, different cellular contexts, titrations, compound series, etc.

However, if such a platform is applied for large sets of perturbation, spanning years of different project stages and various programs, then data analysis, and particularly homogenization, become important. Large scale expression profiling projects such as the Connectivity Map (Lamb, Crawford et al. 2006) and applications presented herein have to contend with day-to-day variation in cellular responses. Indeed, batch effects were previously considered a nuisance that was dealt with using robust rank-based statistics (connectivity score, (Lamb, Crawford et al. 2006)), and via use of biologically motivated data summaries such as Gene Set Enrichment Analysis (Mootha, Lindgren et al. 2003, Subramanian, Tamayo et al. 2005). It is not clear that such nonparametric approaches, which depend on prior knowledge (biological pathways or previous expression experiments), yield the highest possible sensitivity and specificity for downstream analyses.

Herein we present a novel method of representing the expression profiles of the L1000 platform as a short binary barcode. The approach starts by training a deep model that learns to distinguish replicate from nonreplicate profiles. The internal state of the model thus learned demonstrated properties previously ascribed to deep neural network models, namely a hierarchical representation that captures the regularities and underlying structure of the input data (Bengio 2012, Bengio 2013). The internal state represents a robust, abstracted representation of the data, one that captures inherent aspects of the biology such as similarity of compound targets and pleiotropy. We go on to apply this framework to prospectively predict the targets of compounds based on the transformation of their induced expression profiles.

# Methods

## L1000 data generation

Doses for treatment were chosen using the following hierarchy of approaches: for compounds with known biological activity and toxicity, the dose with the highest therapeutic window was selected. For compounds with only biological activity, the $EC_{80}$ was used. For compounds without relevant cell-based assay activity, the highest non-toxic dose was used, up to 20μM. Two cell lines were used: PC3 and ME180. A number of compounds were profiled in both cell lines, the rest were profiled in just one of the two. Most compounds were profiled as biological replicates by performing identical treatments and lysis on separate days, some compounds were only represented by a single instance (the latter contributed only to negative example pairs in the metric learning). Compound treatment time was 6 hours, after which cells were lysed, frozen, and shipped to Genometry for processing. A total of 3699 compounds were screened. Compounds were selected based on being tool bioactive compounds, active compounds for ongoing phenotypic screening programs, and compounds of interest for particular compound optimization programs. Compound profiling was performed in four independent campaigns over the course of 14 months.

## Data processing

Initial data processing was conducted by Genometry using a standard pipeline. Briefly, Luminex intensity measurements were assessed for consistency of relative expression of control genes. Samples passing well and plate level thresholds were summarized by conversion of intensity data to calculated $\log_2$ Genechip-equivalent intensities and normalized based on control gene intensities. Finally, the data on each plate was standardized based on the median and median absolute deviation of the vehicle control samples to calculate z-scores for each gene. These z-scores were the input for the deep metric learning, and were also analyzed directly as a baseline for data interpretation.

## Deep metric learning

A metric learning cost module that takes consecutive pairs of training examples and calculates a distance between them, and thereby a cost, was implemented using the Pylearn2 framework (Goodfellow, Warde-Farley et al. 2013). Internal data representing L1000 data as z-scores of 7573 profiles of 3699 compounds was used. Train/validation/test datasets were made from the initial data by selecting respectively 80, 10, and 10 % of the initial 3.7k compounds. Within a dataset, a sample of pairs of profiles representing biological replicates was used for positive examples (n=40k), and in addition a sample of an equal number of pairs of samples that were not biological replicates was used as negative examples. Hyperparameters (number and type of layers, regularization, dropout), were tuned manually based on accuracy of the model on the validation dataset, and the model depicted in Figure 2 emerged as the best performing. In order to build a model that facilitates application to large scale profiling datasets, it was of interest to bias the model to produce a representation that could be used akin to locality sensitive hashing (Masci, Bronstein et al. 2014), or semantic hashing (Salakhutdinov and Hinton 2009), whereby large collections of profiles can be queried for similar expression profiles without performing global similarity calculations (which become prohibitive as datasets increase in size). Hashing of L1000 data is the subject of ongoing work, and is not explored further here. Nevertheless, it was found a model designed to facilitate hash-based lookup (i.e. having saturated, nearly binary internal representations), gave a higher validation set accuracy than other models tested, so this is the model that was used for downstream analyses.

The model uses an input layer of 978 z-scores, followed by two hidden layers of 400 and 100 units, using a noisy sigmoid nonlinearity:

$$y = \sigma(Wx + b + \mathcal{N}(0, 0.25I)) \quad (1)$$
$$\sigma(z) = (1 + e^{-z})^{-1} \quad (2)$$

Independent Gaussian noise with mean 0 and variance 0.25 ($\mathcal{N}(0, 0.25I)$), is added to the weighted sum of inputs plus bias ($Wx + b$) before applying sigmoid nonlinearity σ. Noise was added to favor saturated (0/1) activations. Dropout (Srivastava, Hinton et al. 2014) ($p$=0.5) and L1 weight decay ($\lambda$=$10^{-5}$-$10^{-6}$) are also used in hidden layers for regularization. Finally, the cost layer uses a rectified loss function with a squared distance margin of 5:

$$c = softplus(1 - y(m - d^2)) \quad (3)$$
$$softplus(x) = ln(1 + e^x) \quad (4)$$
$$y = \begin{cases} 1 \text{ if pair are replicates} \\ -1 \text{ if pair are non-replicates} \end{cases} \quad (5)$$

Here $c$ is the cost, $m$ is the margin, $y$ is the training target (-1 for non-replicate, +1 for replicate), $d^2$ is the squared Euclidean distance between hidden vectors. Theano (Bastien, Lamblin et al. 2012) automatic differentiation is used to backpropagate the cost to the model weights (including regularization) via the pylearn2 (Goodfellow, Warde-Farley et al. 2013) framework. The model parameters are optimized using the RMSprop (Tieleman TH 2012) algorithm for adaptive minibatch stochastic gradient descent.

After training the model, the weights were extracted and used to generate the activation state of the last hidden layer for each input sample. While noise was added during training to regularize the model and encourage saturation, no noise was added during barcode generation in order to yield a deterministic transformation. Perturbation barcodes are generated by thresholding activations of the last hidden layer to create a binary representation. See 'Availability of supporting material' for software.

**Gene set enrichment analysis**

GSEA was conducted using 6034 gene sets derived from experimental gene signatures of up- and down-regulated genes curated by Nextbio (Santa Clara, CA). Enrichment was measured using a Wilcoxon test, and the resulting rank sum test z statistic was used as a score for a given gene set's enrichment in a sample.

**LINCS consortium data**

Data downloaded from the NCBI GEO repository, accession GSE70138. The file GSE70138_Broad_LINCS_Level4_ZSVCINF_mlr12k_n78980x22268_2015-06-30.gct.gz was downloaded in December 2015. This dataset represents 273 compounds tested in 6-point dose response at two time points in 15 cell lines. The level 4 (z-score vs. vehicle control) data was utilized. The first 978 landmark (i.e. measured) gene measurements of the data matrix were analyzed. In order to train the perturbation barcode model on the LINCS data, a subset of the data was selected: 10 and 1 µM treatments at 24 hours. 80% of the data was used for training, and 20% for validation, and the same hyperparameters were used for the model. After the model was trained, the entire dataset was encoded with the perturbation barcodes learned from the training set. Targets for 149 of the compounds were found to be annotated in Chembl (Gaulton, Bellis et al. 2012) or Metabase (Ekins, Bugrim et al. 2005) databases, only those targets affected with potencies <1µM were considered. For each target, distances between profiles derived from distinct compounds sharing the target were compared to a sample not sharing the target via a t statistic. Compound structures were clustered by hierarchical clustering of ECFP4 fingerprints, and gene expression profiles were clustered by the clara (Kaufman and Rousseeuw 2009) algorithm using cluster numbers (k) determined from optimal cuts of hierarchical clustering trees of samples of the datasets. The clusterings were compared via the Adjusted Rand Index.

**Other analyses**

Crossvalidated support vector regression was performed with the caret package (Kuhn 2008), clustering, and correlation analyses were performed in R (Team 2012).

For visualization, data, either z-scores or barcodes, were reduced to two dimensions using the t-distributed stochastic neighbor embedding (t-SNE (Van der Maaten and Hinton 2008)) algorithm (implemented in tsne (Donaldson 2012) package for R). Data was visualized by plotting the t-SNE features in Spotfire (Tibco).

Compound structures were clustered using in-house fragment-based descriptors (Clark 2005), and hierarchical clustering with a threshold of Dice similarity>0.6. Promiscuity was defined based on the frequency with which each compound was considered a 'hit' across HTS assays. Project-specific hit calling was used for each assay.

**Known target annotation**

Merck's chemogenomic database, the Chemical Genetic Interaction Enterprise (CHEMGENIE) was used to annotate compounds with their known targets. CHEMGENIE contains harmonized data from external (e.g., ChEMBL, Metabase, and PDB) sources as well as internal sources (e.g., project team data, kinase profile panels, counterscreen panels, etc.) where compounds are represented as desalted InChIKeys and targets are represented with their Entrez Gene ID and/or UniProt Accession. All target-based dose response data was added to each compound, and the highest affinity target was selected as a representative target for each compound.

**High Throughput Screening Fingerprints (HTS-FP)**

HTS fingerprints were constructed as described in the past (Petrone, Simms et al. 2012). Briefly, the z-scores of primary HTS screens at Merck were calculated for all screens where > 1e6 compounds had been screened (344 total screens). The z-score for each screen was then stored as a vector for each compound. Any z-scores > 20 or <-20 was assigned a value of 20 or -20, respectively, so that artifactual values would not impact Pearson correlation calculations as significantly. The Pearson correlation of HTS-FPs was calculated between pairs of compounds by calculated the correlation of z-scores for assays that were in common between them (all other z-scores were ignored). 3471 compounds possessed HTS-FPs and were used in this analysis.

**Activity prediction**

t-SNE visualizations of z-scores representing the normalized L1000 data, and also of the 100-dimensional (nonthresholded) perturbation barcodes derived from the metric learning network were generated. The locations of known EGFR/MEK/MAPK inhibitors on these maps were plotted, and points (i.e. representing compound treatments), that were surrounded by these known actives were selected for testing from each of the two maps. Separately, in order to select actives from the full 978-dimensional z-score or 100-dimensional barcode space, the ten nearest neighbors (by Euclidean distance) of each of the seed compounds were identified, and available compounds were tested for activity. Compounds were tested in dose titration assay in 1536 well plate format using a previously optimized reporter gene assay.

Cells (CellSensor ME-180 AP-1-bla, Life Technologies) were plated in the manufacturer's recommended assay medium at 3000 cells/well in 9μl in black/clear tissue culture treated 1536 well plates (Greiner), and allowed to adhere overnight. Compounds were added from DMSO serial dilution plates (50μM maximum assay concentration, 8-point, 3-fold dilutions), using a 50nl pintool (GNF Systems), and the cells incubated 30 min in a tissue culture incubator. The cells were then stimulated with Epidermal Growth Factor (EGF, Life Technologies) at 10ng/ml final concentration by adding 1μl 10x stock, and allowed to respond for 5 hours in a tissue culture incubator. Beta-lactamase detection reagents (ToxBlazer, Life Technologies), were added per manufacturer's instructions (2μl 6x mix), plates were incubated 2 hours at room temperature, and read using a bottom-reading multimode reader (Pherastar, BMG). Data was normalized to no-stimulation (100% inhibition), and stimulation + DMSO (0% inhibition), controls, and percent inhibition was plotted along with logistic regression curve fits using Spotfire (Tibco). Compounds were classified as active if they exceeded 50% inhibition of reporter activity without toxicity at concentrations below 10μM.

## Results & Discussion

### 1. Transforming gene expression profiles

We first introduce the data generation process and highlight the need for improved analyses. The disparity between replicate profiles motivates the development of a metric learning approach that improves upon the current state of the art for this type of data analysis.

*High throughput expression profiling*

The experimental approach used in this work is depicted in Figure 1A. Cells can be treated, lysed, and measured in medium throughput (384-well) formats. The use of commodity consumables, reagents, and technology permits relatively low cost profiling of hundreds to thousands of compound-treated samples at a time (Peck, Crawford et al. 2006). Reader intensity measurements are normalized using control samples and genes. Next, each perturbation's expression is compared to an internal (within-batch) negative control. Twelve to sixteen vehicle (DMSO) controls are measured on each 384-well assay plate, and the expression of the remainder of samples is standardized to the average (median), and scale of variation (median absolute deviation), of these vehicle controls to yield robust z-scores, henceforth referred to as z-scores for brevity.

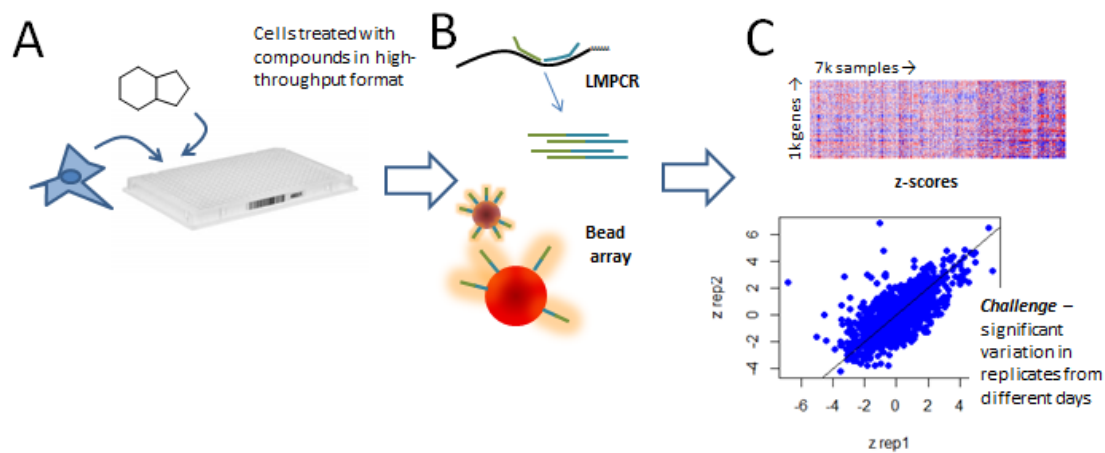

**Figure 1.** Experimental setup and architecture of the deep model used. **A.** Cells treated with compounds in 384-well plates. **B**. Cell lysate used for ligation mediated PCR with gene-specific probe pairs, and the gene expression measured using an optically addressed bead array technology. **C.** Raw intensity is normalized and converted to relative expression changes versus control (z-scores) on a plate-wise basis. Variability observed between biological replicates.

Large scale gene expression profiling is known to have significant 'batch' effects that confound interpretation (Leek, Scharpf et al. 2010). These effects are a result of differences in cell response from day to day. The effects can be controlled for by performing replicate experiments, but the level of replication required to eliminate the batch effects is not practical for high throughput experimentation, which requires knowledge to be extracted from 1-2 replicates. Additional approaches to reduce the influence of experimental batch include normalization of data within a batch such that all perturbations are compared to an internal negative control in the same batch. In spite of measures to control batch effects, it is nevertheless observed that there is variability in cell response that makes it difficult to interpret data obtained on different days (Leek, Scharpf et al.). As an example, for the dataset described below, we can assess the similarity of biological replicates' gene expression profiles using Euclidean distance of the normalized expression changes *versus* control. Looking at the similarity of each sample profile to every other in the dataset, the median rank of a sample's biological replicate is approximately the $3^{rd}$ percentile. While this ranking can be considered an enrichment (versus a null hypothesis of ~$50^{th}$ percentile), the $3^{rd}$ percentile in our dataset nevertheless implies that >200 treatment profiles are more similar to a given treatment than a sample that was treated identically on a different day.

As mentioned above, the z-score data is confounded by batch effects that reduce the apparent similarity of expression responses deriving from identical treatments. In order to improve on the performance of the z-scores, another strategy that was explored is that of gene set enrichment analysis (GSEA, (Subramanian, Tamayo et al. 2005) ). GSEA is a technique used to extract interpretable information from expression profiles, and is useful for placing the up- and down-regulation of genes into a biological context (e.g. biological pathway, disease state). In addition, GSEA profiles (i.e. enrichment scores of a sample across many different gene sets), can be thought of as a means of averaging, and potentially making more robust, the expression profiles by using biological context as a prior. Thus, it might be expected (Edelman, Porrello et al. 2006) that GSEA profiles might be more reproducible, and more predictive, than raw expression data. Indeed, when one looks at the correlation of pairs of samples treated with the same compound, one sees that GSEA profiles show higher concordance than z-score profiles day-to-day and cell-to-cell (Supplementary Figure 1). In the previoiusly mentioned measure of average rank of replicates, GSEA yields a median rank of ~1%, with ~70 profiles more similar to a given sample than its biological replicate.

*Deep metric learning*

Since there is room for improvement over both of the previously described methods, we sought to evaluate alternate strategies to increase the sensitivity and specificity of interpretations and predictions derived from relatively large data sets that we are currently generating on the L1000 platform. In particular, as a first step, we sought to increase the concordance of samples that should exhibit the same phenotypic response, while separating them from those that should be distinct. Using the similarity of biological replicates as a gold standard, the goal was to recast the data in a way that maximizes the similarity of replicates in contrast to non-replicate samples. This can be thought of as a metric learning problem, whereby one learns a new measure of distance or similarity between points, one that distinguished replicates from non-replicates. Initial experimentation with linear and kernel methods (Bellet, Habrard et al. 2013) failed to yield significant improvements over the raw data, thus deep neural network models were next evaluated.

Deep learning is an extension of decades-old artificial neural networks, and has recently shown impressive performance on a number of image and speech recognition tasks (Hinton, Deng et al. 2012, Krizhevsky, Sutskever et al. 2012). As the properties and capabilities of these models have begun to be elucidated, it is becoming more feasible and reproducible to create highly accurate multilayer neural network models. In addition, advances in model design have been complemented by increased computational power and larger datasets that enable the training of comprehensive, expressive models, whose generalization performance is facilitated by regularization (Bengio 2012, Bengio 2013). In addition to providing a flexible model that is able to fit complicated datasets, deep networks have the additional advantage of learning a hierarchical representation of the data, whereby lower layers of the model learn to represent the fine-grained detail of the input data, and higher layers represent increasing layers of abstraction. Thus, inspecting the higher layers of a model has the potential of revealing the underlying high-level structure of the data, with reduced sensitivity to noise and contextual details. We thus explored a multilayered network with the dual goal of developing a suitable distance metric, and also developing a robust data representation for downstream analyses.

Most machine learning methods consider a single input sample at a time, though each sample is generally represented by multiple variables, or features. The goal of this effort was to learn not about individual samples, but about the comparison of pairs of inputs in order to learn a new metric that better captures what is known about the data. Thus, a method was chosen that permits comparative learning in a neural network framework. We implemented a version of a siamese neural network (Liu 2013, Hu, Lu et al. 2014) to perform metric learning on a dataset of L1000 data (Figure 2).

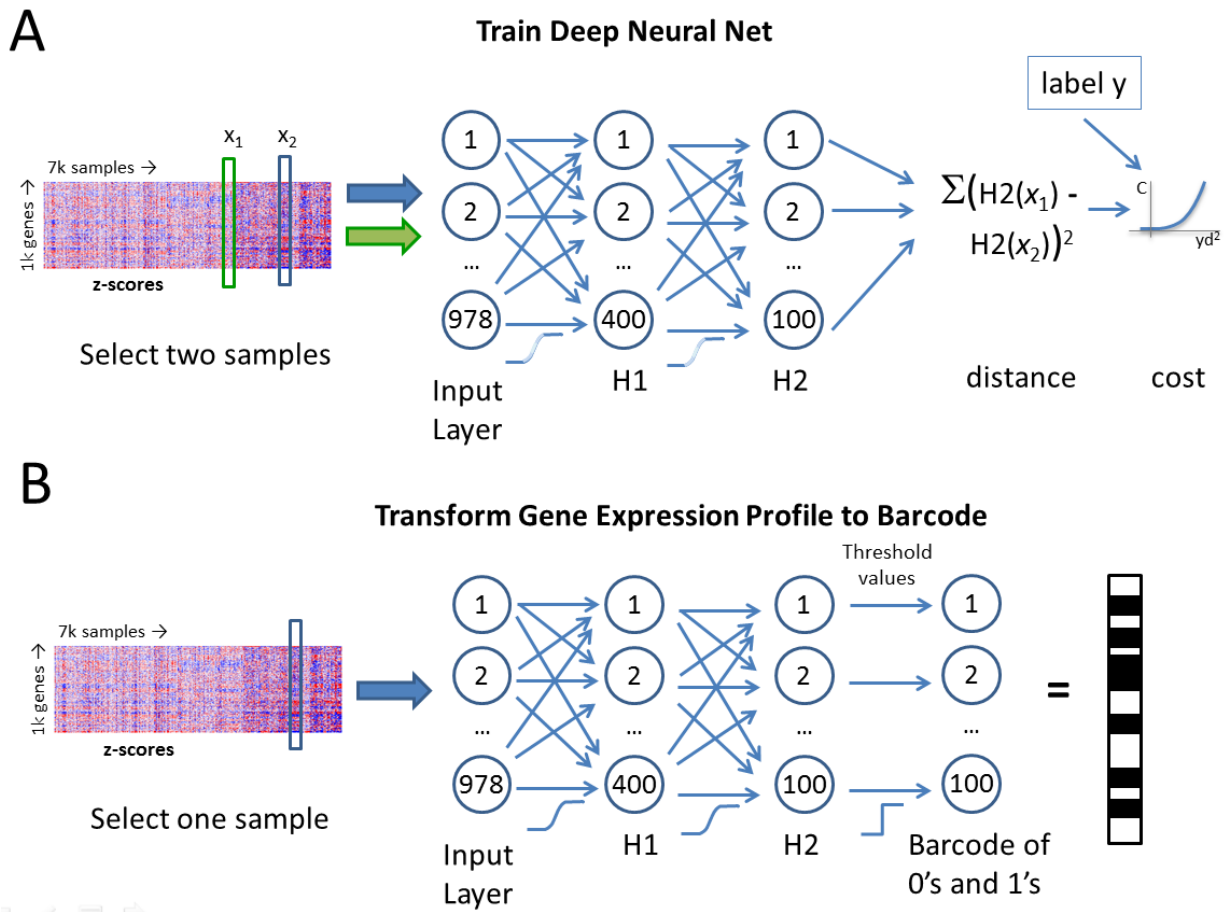

**Figure 2. A:** Metric learning network: a pair of 978-element z-score vectors is input to the network as adjacent vectors. Data is transformed through two layers (400 followed by 100 units), of nonlinearities (noisy sigmoid activation functions). The activations of the second hidden layer ($H2(x_1)$ and $H2(x_2)$) are combined in the cost layer by calculating a Euclidean distance between the two representations. The margin cost is calculated based on the -1/1 (non-replicate/replicate indicator) target and the squared distance. **B.** Once the model is trained, expression profiles are converted to barcodes by passing them through the first two (now noiseless sigmoid) hidden layers and thresholding the activation of the second hidden layer to yield perturbation barcodes.

A number of model architecture properties (also known as hyperparameters) need to be set prior to the optimization of model weights of a neural network by gradient descent. These include number of layers, layer sizes, activations of layers, regularization types and weights, and cost type and any parameters. We undertook a greedy manual search of hyperparemters, optimizing the validation set accuracy for predicting replicate vs. nonreplicate pairs. Among over 60 models tested, we settled on the architecture described below since it provided the best performance on validation data.

The final model that emerged from optimization of the neural network architecture performs the following: a pair of 978-dimensional z-score vectors representing two different expression profiles is given to the network as input, the data is transformed by two layers of noisy sigmoid layers, and then the representations of the members of the pair are compared to each other by calculating a Euclidean distance. A margin loss is calculated based on the distance *versus* the known target status of the pair (i.e. replicates or not replicates). The cost is used to train the network via backpropagation of the cost gradients (Rumelhart, Hinton et al. 1988). While siamese networks have been previously depicted as a linked pair of networks transforming the pair of inputs for comparison (see (Liu 2013, Hu, Lu et al. 2014)), in practice we found it simpler to implement the siamese network model as a single network that processes pairs of samples as adjacent inputs (Figure 2). In this architecture, the paired input is provided as two consecutive vectors to one network that processes the elements of the pair through the aforementioned hidden layers and then calculates the cost.

The network was trained on 80% of the compounds, using a validation held-out set of 10% to select model architecture and detailed configuration (hyperparameters), and a final test set of 10% was used to estimate generalization performance of the selected model. The trained model was able to correctly categorize 97% of test pairs as replicates/non-replicates ($F_1 = 0.87$, (Van Rijsbergen 1979) ). The deep model performed significantly better than a linear metric learning model (a single hidden layer model with linear activation function, otherwise same as deep network), which achieves 74% accuracy ($F_1 = 0.47$). As a baseline, a model that randomly samples from the empirical distribution of replicates/non-replicates would have an accuracy of 56% ($F_1 = 0.33$).

Having demonstrated that it is possible to recast the data in a way that captures the similarity of identically treated samples, we set out to determine if the learned model can be used to improve interpretability of large scale expression profiling campaigns. Since the model can separate replicates from nonreplicates, and since the model output is a function of a simple Euclidean distance calculated on the internal representation (hidden layer activations), we reasoned that this internal representation captures the discriminatory power learned by the model. We explored whether this internal representation has additional useful properties, as for example do the internal word representations of skip-gram language models (Mikolov, Chen et al. 2013). The internal representation was extracted from the learned model by using the activation of 100-dimensional second hidden layer as the learned feature vector of each data point (Figure 2B). This representation is a short, fixed-length, almost binary representation that captures expression profile changes. In order to facilitate computation of similarities and hash-based lookup, the almost-binary representation (~95% of activations >0.9 or < 0.1), was thresholded to be exactly binary (i.e. values ≥ 0.5 were set to 1, rest to 0). This binary representation of cellular gene expression profile changes in response to treatments is referred to as a 'perturbation barcode.'

Thus the model yields a new encoding of the original data, a representation that was designed only to increase the similarity of biological replicates relative to non-replicates. Information is necessarily lost in reducing 978-dimensional continuous data to 100-dimensional binary representations. The question that remains is whether biologically interesting aspects of the data are retained in the simplified encoding.

Thus, we evaluated whether the model described herein learns useful, generalizable aspects related to compound effects on cells as a byproduct of learning to identify replicates. The perturbation barcodes were assessed for their ability to identify similarities between compounds in a variety of contexts (Table 1). Furthermore, we directly compared the performance of the barcodes to the minimally processed z-score data and the gene set enrichment score profiles.

The median rank (out of 7573) of each sample's biological replicate is shown in Table 1. For each sample, the similarity of all other samples' profiles is ranked, and the median rank across samples of the biological replicate is reported. For the z-score data this value is 225, indicating that, on average, there are 223 profiles more similar to a given sample (itself ranked #1), than the sample's biological replicate. The median rank for GSEA score data is 72, and for perturbation barcodes it is 24. The result on the perturbation barcodes demonstrates that the training objective was met: samples that are replicates are more similar under the features derived from the metric learning than they are in the original data space.

| metric | z-score | GSEA | perturbation barcode |
|---|---|---|---|
| Median rank of replicates (of 7573) | 225 | 72 | **24** |
| Distance by shared target, t statistic | -1 | -38 | **-43** |
| Structural clustering overlap with expression clustering | 0.01 | 0.03 | **0.17** |
| Correlation of HTS profiles with expression | 0.04 | 0.02 | **0.12** |
| Promiscuity prediction by SVR, $R^2$ | 0.21 | 0.16 | **0.34** |

**Table 1.** Performance of learned perturbation barcodes compared to z-scores and GSEA scores. **Row 1:** For each sample with replicates in the dataset, profiles are ranked based on Euclidean distances calculated from the various representations, and the median value of the replicates' ranks across samples is reported. **Row 2:** Distance of pairs of profiles of compounds that share a target annotation compared to those that do not. Significance of difference in mean distance measured with a t statistic. For reference, a permutation analysis of the target labels in the barcode dataset yielded a minimum t statistic of -5.8 from 100 random permutations ($p<0.01$). **Row 3:** Compounds clustered based on structure and on the expression profiles they induce. The overlap of the structural and expression clustering is measured by the Adjusted Rand Index on a 0-1 scale. For reference, a permutation analysis of the cluster labels in the barcode dataset yielded a maximum ARI of 0.002 from 100 random permutations ($p<0.01$). **Row 4:** Similarity of phenotypic profiles measured either by activity across HTS assays, or by induced expression changes. The correlation of each expression measure to the HTS fingerprint data is shown. For reference, a permutation analysis of the sample labels in the barcode dataset yielded a maximum correlation of 0.001 from 100 random permutations ($p<0.01$). **Row 5:** A support vector regression model was trained using the various expression features to predict compound promiscuity (fraction HTS screens in which a compound is active). Crossvalidation performance is measured using $R^2$ of predicted *vs.* observed promiscuity. The standard error of barcode $R^2$ values was 0.06 ($p<0.01$).

## 2. Comparison of Perturbation Barcode with State of the Art

Next, one might ask whether a model trained to recognize samples treated with the same compound would also retain additional relationships related to biological function and chemical similarity. In order to avoid re-discovering the similarity of replicates, for this and subsequent comparisons, replicates were resolved to a single representation by averaging the hidden layer 2 activations of the replicate samples before thresholding to barcodes, and analogously, z-scores and GSEA results represent the average of replicates.

We tested the ability of the different representations to recognize different compounds targeting the same cellular factor as having similar profiles. Due to compound-specific polypharmacologies, one shouldn't expect any pair of distinct compounds to have identical profiles, but one should expect compounds sharing a target to have profiles that are more similar, on average, than pairs of compounds not sharing a molecular target. We looked at those compounds for which target annotation was available (1297 compounds having annotated targets modulated with potency < 1μM), and asked if compounds that share a target are more similar to each other in induced gene expression profiles than compounds that do not. Using a t statistic for the difference of mean distances of shared *versus* unshared targets, the comparison on the z-score data is not significant, with a t statistic of -1. The GSEA-summarized data is more significant with t~-38, while the barcode data shows the highest preferential similarity of profiles of compounds sharing a target annotation: t~-43. This indicates that the latter two methods are able to encode the expression profiles of compounds in a way that emphasizes biological similarity.

A well-studied approach to categorize compounds is by similarity of chemical structure, with one underlying assumption being that similarity of structure is related to similarity in function. Thus molecules are clustered using descriptors designed to allow molecules considered similar by a trained chemist to be grouped together, to the exclusion of dissimilar molecules (Brown and Martin 1996). Analogously, expression profiles can be clustered to find sets of genes or of samples that are more similar to each other than to other instances, in an attempt to identify cohesive functional groupings (Alizadeh, Eisen et al. 2000). One might expect a significant but imperfect overlap between compound structural clusters and those derived from expression profiling. The expectation of significant overlap comes from the fact that compounds with shared scaffolds or similarly arranged functional groups would be more likely to target the same cellular factors and thus yield similar expression profiles. Imperfect overlap would be expected because small changes in structure can lead to significant changes in activity, a phenomenon studied in structure-activity relationships of chemical series (Stumpfe and Bajorath 2012). In order to compare overlap of structure and phenotype, a comparison of the clusterings of the two types of data was performed. Compound structures were clustered using hierarchical clustering of fragment-based descriptors with a minimum similarity criterion to define clusters. Expression data was clustered using hierarchical clustering followed by a dynamic tree cutting (Langfelder, Zhang et al. 2008) approach to generate discrete, well separated-clusters. The overlap of clusterings is measured with the Adjusted Rand Index (Hubert and Arabie 1985) which varies from 1.0 for perfect overlap, to 0 for overlap expected by chance. The z-score data shows little overlap of expression clustering with the structural clusters (1%, Table 1), and the GSEA-based clusters modesty better (3%). In contrast, the perturbation barcode clusters have significantly higher overlap with structural clusters than z-scores (17%). These results are not sensitive to the clustering algorithm nor to compound structural descriptors, as similar observations were obtained with affinity propagation clustering and atom pair descriptors (Sheridan, Miller et al. 1996) (not shown).

Next we set out to determine whether the barcode reveals insight into biological function. In order to compare two different measures of biological activity, the profile of 3471 compounds across multiple High Throughput Screens (HTS) was compared to the L1000 expression profiles. The activity profiles across HTS assays could be thought of a biological fingerprint, as it represents the activity of the compound in a multitude of disparate assays, both cell-based and biochemical. Such profiles do not evenly span all potential biological responses since they opportunistically derived, depending on the history of each compound and the organization performing the HTS campaigns. However, in some ways the HTS profiles are more relevant to drug discovery since many of the constituent assays directly measure the molecular activities of interest rather than induced transcriptional programs. The correlation matrix of HTS fingerprints (HTS-FP, (Petrone, Simms et al. 2012)) of the compounds was compared to the correlation matrices of z-scores, GSEA scores, and perturbation barcodes. The correlation of the triangular correlation matrices (Mantel statistic) was used as a measure of how similar the HTS activity profiles are to z-score, GSEA, and perturbation barcodes. The perturbation barcodes show more than three times the similarity to HTS profiles compared the other two scores (Table 1).

Finally, it was of interest to explore if there is a difference in the ability of the three data representations to predict a high-level property of compounds, namely their promiscuity. The promiscuity of a compound is defined here as the fraction of the HTS assays in which the compound was scored as active, ranging from 0 to 0.5 in our dataset (median 0.04, 3836 compounds with data). Promiscuity of compounds is of interest in drug discovery, wherein modulators that are both potent and specific (i.e. non-promiscuous) are generally sought. One way to ascertain if promiscuity information is present in the different expression profile representations is to assess the ability of each representation to generate a model that can quantitatively predict this property for samples not in the training set. Predictive support vector regression models were trained on each data representation, and the average cross-validation coefficient of determination ($R^2$) was used as measure of the ability of the models to predict properties of compounds on the basis of the profiling data (Table 1). Given that the perturbation barcodes are derived solely from the z-score data, and that they are in some ways a more impoverished representation of the original data (978 dimensional real values to 100-dimensional binary values), it was surprising to find that models trained on the barcodes are able to explain ~60% more promiscuity compared to the z-scores.

To test whether the feature mapping can be applied to other datasets, we trained an identical model architecture on a sample of L1000 data generated by the LINCS consortium (Duan, Flynn et al. 2014). For the comparisons we could make, we found the same benefits of the perturbation barcode versus the raw data. Table 2 shows that the learned barcode brought biological replicates closer to each other (median in top 1% most similar profiles *versus* top 27% for z scores). Furthermore, average profiles of distinct compounds sharing a target are significantly closer to each other in the perturbation barcode space than in the original space: for each annotated target of compounds in the dataset, the distances of compounds sharing *versus* not sharing the target were compared via a t-statistic, and the mean of these statistics was found to be non-significantly higher (+1) for the z-score data but significantly smaller (-37) for the barcodes (Table 2). Finally, although this dataset contained a relatively small number of structurally distinct compounds, we could still detect a greater than 2-fold increase in correspondence between clustering of transcriptional profiles and structural descriptors in the perturbation barcodes versus the z-scores (Table 2).

| metric | z-score | perturbation barcode |
| --- | --- | --- |
| Median rank of replicates (of 79890) | 21496 | **649** |
| Distance by shared target, mean t statistic | 1 | **-37** |
| Structural clustering overlap with expression clustering | 0.004 | **0.010** |

Table 2

### 3. Prospective Activity prediction

In order to compare the ability of the learned features to capture the biological activity of tested perturbations in an exploratory setting, the predictive ability of the perturbation barcodes was tested. We compared the ability of the perturbation barcodes and z-scores to retrieve molecules with similar targets in both a two-dimensional visualization, and by nearest neighbor searches in the respective native data spaces. t-SNE (Van der Maaten and Hinton 2008) is a dimensionality reduction technique used for visualization in large datasets. The transformation provided by t-SNE is not quantitative and does not preserve all aspects of the original data space, but in a 100-1000-dimensional data space, t-SNE is the best system we have found for exploring such data sets. The predictive fidelity of the t-SNE map derived from expression z-scores was compared to that of a t-SNE map derived from the perturbation barcodes in order to ascertain how much information one can hope to glean by studying the visualization of a large dataset (Figure 3A - C). This use case was studied because it is representative of a common query one may want to pose of a dataset: given knowledge about the molecular mechanism of action of some compounds, can we expect compounds nearby to them in a visualization to share similar mechanisms? Three known inhibitors of the EGF/MAPK pathway that were tightly clustered in both maps were chosen as seeds, and 31 unknown compounds found nearby the known MAPK pathway actives in either or both t-SNE maps were tested for their ability to block signaling from EGF to AP1 using a reporter gene assay. The compounds selected in this way were structurally diverse, with only two of the compounds sharing the same structural cluster. Figure 3E shows that there were two unknown compounds that were identified in the z-score t-SNE maps and were confirmed to be active in MAPK signaling. By comparison, the barcode map revealed the same two and four additional novel unknowns (Figure 3F). Negative control compounds selected to be far away from the MAPK tools in both maps showed no specific AP-1 reporter activity (Supplementary figure 2). One compound that was a known pathway inhibitor did not confirm in this assay as it showed high potency ($IC_{50}$~35nM), but low efficacy ($E_{max}$~41%), thus not meeting the activity threshold. The additional actives discovered only in the barcode map demonstrate that the visualization derived from the barcodes has increased sensitivity to capture biological activities *versus* the raw data-derived visualization. Significantly, none of the newly identified AP-1 reporter inhibitors were structurally similar to the known actives (Dice similarity less than 0.6).

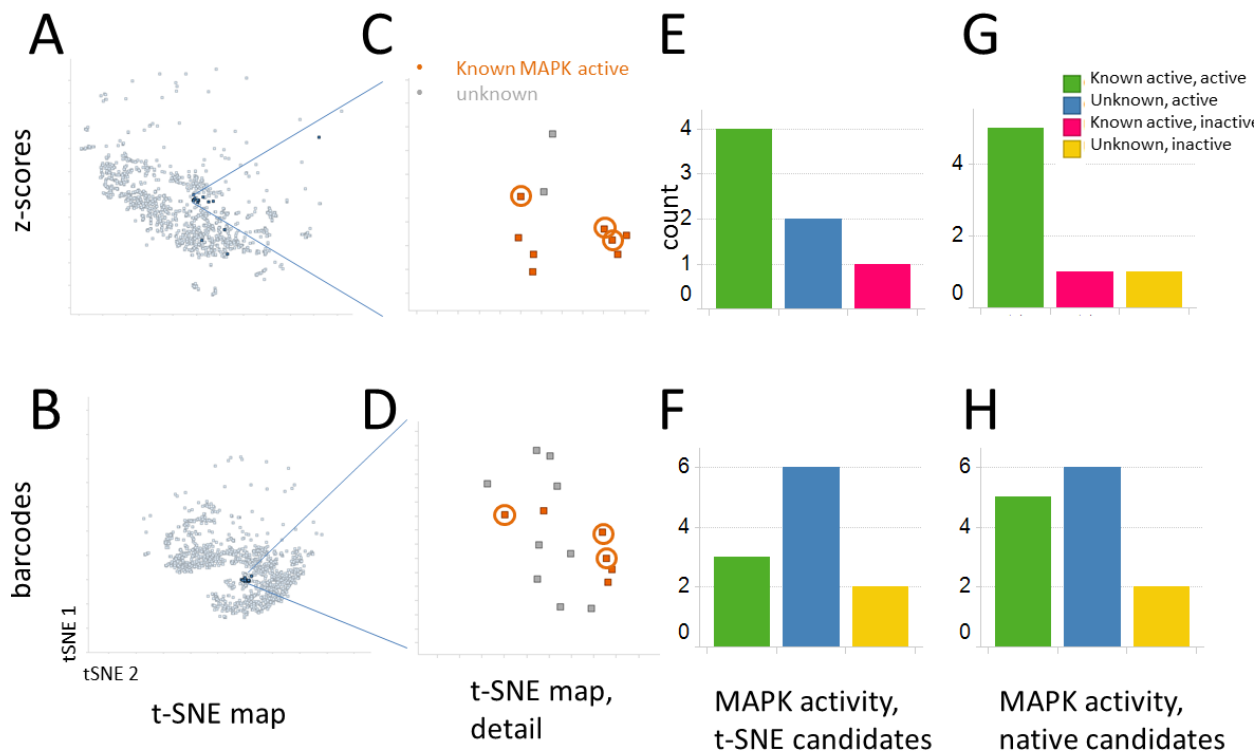

**Figure 3.** Visualizations of the data based on z-scores or perturbation barcodes were examined to select candidate compounds in the phenotypic neighborhood of a series of known MAPK pathway inhibitors. **A - D:** t-SNE maps of the data, z-scores on top, perturbation barcode maps on the bottom. **A, B:** the entire dataset is shown with the tested compounds in dark blue. **C,D:** The neighborhood of the query MAPK pathway inhibitor compounds (orange) is shown. Common MAPK tools used for nearest neighbor analysis are circled. **E,F:** Results of AP-1 reporter assays. Known MAPK actives are distinguished from unknowns predicted to be active in C,D. **G,H:** Rather than selecting neighbors of seed MAPK tool compounds in the t-SNE map, nearest neighbors in the native datasets were selected and tested in the AP-1 reporter assay. Key as in E,F. See supplementary figure 2 for breakdown by categories, including overlaps.

Finally, we looked at the neighborhood of the same known MAPK actives in the original data space for both the z-scores (978-dimensional) and the barcodes (100-dimensional) rather than the reduced 2-dimensional t-SNE space. Nearest neighbors were chosen by Euclidean distance and tested in the reporter assay as above. Figure 3G & 3H show that a larger number and fraction of unknowns identified as candidates from the barcode data were confirmed to have AP-1 reporter activity compared to the z-scores.

## Conclusions

We present here a straightforward way to extract features from gene expression data that are in many ways more expressive and robust than the original gene expression changes. While training of a deep network is a somewhat specialized process, once an adequate model is trained, additional samples can efficiently be converted to perturbation barcodes with no (or little), further optimization needed, as evidenced by the simple transfer of the model to external (LINCS) data.

Others have proposed novel methods for analyzing L1000 data. In particular, Liu et al (Liu, Su et al. 2015) presented a pipeline that incorporates raw intensity processing and gene assignment through gene set enrichment and production of features informed by protein interaction data. While this type of biological annotation is potentially useful, our proposed perturbation barcode is self-contained, and does not rely on noisy and incomplete biological databases for utility. In addition, the improvements in performance of the learned features compares favorably with other approaches used in the field, including GSEA. Finally, it was shown that the perturbation barcodes can be used to meaningfully predict pathway modulation activity of compounds prospectively. This ability is of value in uncovering unknown modes of action (either primary or secondary) of compounds of interest, and the fact that it can be read directly from a visualization, shows the potential of the approach to simplify and enhance hypothesis generation from big data.

We demonstrate that similarity in the barcode space is indicative of more similarity in compound target and compound structure, activity across biological assays, and predictivity of biological action. Furthermore, the somewhat unexpected observation that models of general compound properties like promiscuity can be better built using the learned features than the data that the features were learned from, indicates that there are positive side effects of the compressive data encoding. We attribute this phenomenon to the denoising property of the learned features, or equivalently, extraction of robust underlying biological factors from the corrupted versions observed in experimentation (due to measurement error, batch effects, and stochastic variation in response).

We anticipate that elaborations of approaches such as this one will be fruitful for other applications in biological and chemical domains, as they have been in artificial intelligence. A more ambitious goal of algorithmic design of molecules may be based on the combination of phenotypic information, quantitative structure-activity relationships, and pharmacokinetic/pharmacodynamic models with generative models for chemical structures. In the meantime, the direct application of the described metric learning-based representation technique to other high volume, high dimensional data has the potential to significantly reduce the effects of noise and to improve interpretability and the quality of generated hypotheses.

## Availability of supporting materials

Code for performing deep metric learning and a demonstration of the analysis on the LINCS data are available at *https://github.com/matudor/siamese* .

## Competing interests

The authors declare that they have no competing interests

# Supplementary Figures

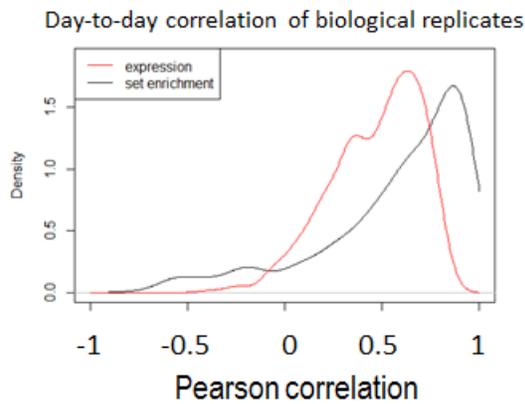 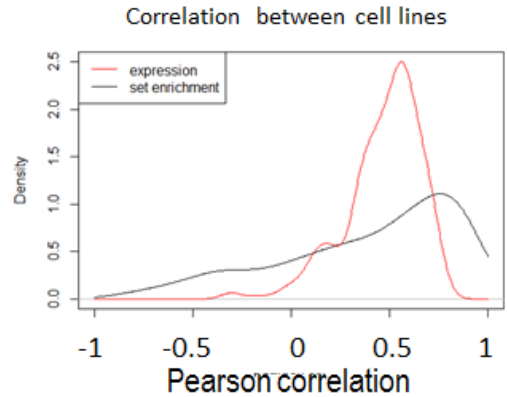

**Supplementary Figure 1.** Correlation of profiles of samples treated with the same compound and dose on different days **A**, or in different cell types **B**. The red line indicates z-scores of gene expression changes, and the black line indicates gene set enrichment scores derived from the z-scores and a library of experimentally derived gene sets. The distribution is smoothed using a Gaussian kernel density estimate.

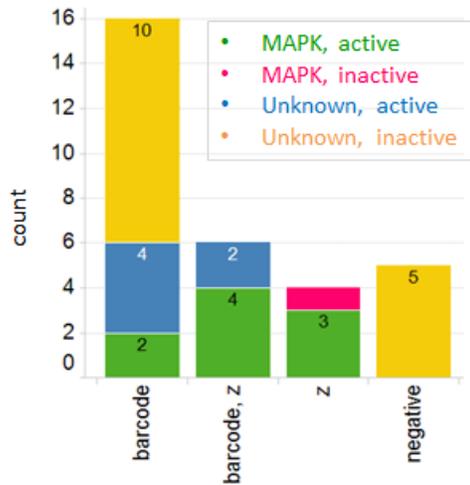 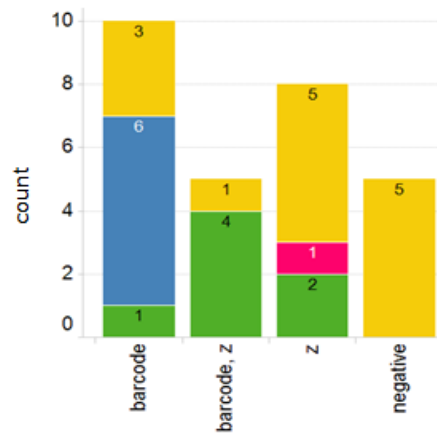

**Supplementary Figure 2.** Activity prediction results. **A**, candidates selected based on t-SNE maps of z-scores, barcodes, or both, are shown along with activity in the EGF/AP1 reporter assay. **B**, candidates selected based on the native datasets (978-dimensional) z-scores, (100-dimensional) barcodes, or both, are shown along with the observed activity in the EGF/AP1 reporter assay.